\magnification=\magstep1
\vsize=22 true cm
\hsize=15 true cm
\baselineskip 24 true pt
\centerline{Quark Matter in Strong Magnetic Field}
\bigskip\medskip
\centerline{Somenath Chakrabarty}
\bigskip
\centerline{\footnote\dag{Permanent Address}Department of Physics, University of Kalyani}
\centerline{Dist.:Nadia, West Bengal}
\centerline{India 741 235}
\centerline{e-mail:somenath@klyuniv.ernet.in}
\centerline{and}
\centerline{Inter-University Centre for Astronomy and Astrophysics}
\centerline{Post Bag 4, Ganeshkhind}
\centerline{Pune, India 411 007}
\bigskip\bigskip\bigskip
\centerline{\bf{Abstract}}

\noindent  The  effect  of strong magnetic field on the stability
and gross properties of bulk as well as quasi-bulk  quark  matter
has  been investigated using the conventional MIT bag model. Both
the Landau diamagnetism and the paramagnetism of quark matter has
been studied. How the quark hadron phase transition  is  affected
by  the  presence of
strong  magnetic  field  has also been investigated. The
equation of state of strange quark matter  changes  significantly
in   strong  magnetic  field.  It  is  also  shown  that the thermal
nucleation of  quark  bubbles  in  compact  metastable  state  of
neutron  matter  is  completely  forbidden  in  presence of strong
magnetic field.
\vfil
\noindent PACS NO.: 25.75+u, 97.60Jd, 71.30+h, 75.20-g
\eject
\noindent {\bf{1. ~Introduction}}

\medskip

It is generally believed that if the density of neutron matter at
the  core  of  a  neutron star exceeds a few times normal nuclear
matter density, a deconfining transition to quark matter may take
place. Whatever be the order of this transition, will  convert  a
normal  neutron star to a hybrid star with an infinite cluster of
quark matter at the core and a crust of neutron  matter.  If  the
speculation  of Witten [1] that a flavor symmetric strange quark
matter  (SQM)  is  the  true  ground  state  of  matter  at  zero
temperature  and pressure is correct, then there is a possibility
that the whole neutron star will be converted to a star of SQM, in
which strange quarks are produced through the weak processes  [2]
and  ultimately  giving  rise to a dynamical chemical equilibrium
among the constituents. The stability of such bulk and quasi-bulk
objects have been thoroughly studied using MIT bag model  [3-6]
as  well  as other confinement models, e.g. we have studied using
{\it{Dynamical Density  Dependent  Quark  Mass}}  [2,7-9]  (D3QM)
model  of  confinement, both at zero and finite temperatures. In this
context we should mention that the results of MIT bag model does not
agree with the recent lattice Monte Carlo results. The latter predicts
a weak first order or a second order QCD phase transition, whereas
according to MIT bag model it is a strong first oredr transition. On the
other hand, we have seen that the predictions of D3QM model more or less
agree with the recent lattice Monte Carlo results. Since we do not know
how to incorporate the effect of strong magnetic field in lattice gauge
calculations, we shall concentrate only on the MIT bag and D3QM models.
To see the effect of strong magnetic field on second order quark-hadron
phase transition we shall assume a metal-insulator type of transition in
presence of strong magnetic field. The
stability and gross properties of hybrid and strange  stars  have
also  been investigated [10-12]. Although the gross properties of
such self-gravitating objects do not differ significantly from  a
pure neutron star (except the M-R diagram), the presence of quark
phase  at  the  core  may cause rapid cooling of the star and can
also produce significant gamma ray bursts.  The  other  important
aspect  is  the  superconducting nature of the quark matter core.
Since quarks carry electric charges, unlike neutron matter  which
exhibits  superfluidity  in a neutron star, this particular phase
as a whole behave like a  superconducting  fluid  (and  possibly
type  I;  of  course  if  there  are  very few protons present in
neutron matter, they may form  Cooper  pairs  and  give  rise  to
superconducting  zones). As a result in such objects the magnetic
field lines mainly pass through the crust region  (if  any),  the
consequence  of  which  is  the rapid Ohmic decay of the magnetic
field. Therefore the surface magnetic field for such objects will
be low enough compared to the corresponding pure neutron star.

The other interesting part of such objects, which must be studied
is  the  time  of  quark matter nucleation. It is still not clear
whether  such  transition  takes  place  immediately  after   the
supernova  explosion  or  much  latter  [13]. If it occurs (if at all) almost
simultaneously with the explosion, then what  is  the  effect  of
strong  magnetic  field  present  in  the  neutron  star  on  the
(first or second order)
quark-hadron phase transition? It is  also  interesting  to  study  the
effect of strong magnetic field on the equation of state of quark
matter.

Now from  the observed features in the spectra of pulsating accreting
neutron star in binary system, the strength of  surface  magnetic
field  of  a  neutron star is found to be $\sim 10^{12}$G. In the
interior of a newly born neutron star,
it  propably  reaches  $\sim  10^{18}$G  [14].  It  is
therefore  advisable to study the effect of strong magnetic field
on SQM, which we  expect to present at the core of a compact
neutron  star. As is well known, the energy of a charged particle
changes significantly in the quantum limit if the magnetic  field
strength  is  equal  to  or  greater  than  some  critical  value
$B_m^{(c)}=m_i^2c^3/(q_i\hbar)$ in Gauss; where $m_i$  and  $q_i$
are respectively the mass and charge (absolute value) of the particle
(e.g.  $q_i=2e/3$ for $u$-quark, $e/3$ for $d$ and $s$ quarks and
$e$ for electron, here $e=\mid e\mid$ is the  absolute  value  of
electronic  charge), $\hbar$ and $c$ are respectively the reduced
Planck constant and velocity of light, both of which  along  with
the  Boltzmann constant $k_B$ are taken to be unity in our choice
of units. For an electron of mass $0.5$MeV, the strength of  this
critical  field  as  mentioned  above  is $B_m^{(c)(e)}\approx 4.4\times
10^{13}$G, whereas for a light quark of current mass $5$MeV, this
particular value becomes $= 10^2\times B_m^{(c)(e)}$,  on  the
other  hand  for  $s$-quark of current mass $150$MeV, it is $\sim
10^{20}$G, which is too high to realize at the core of a  neutron
star.  Therefore  the  quantum  mechanical effect of neutron star
magnetic field on $s$-quark may be ignored. The critical magnetic
field as defined above is  the  typical  strength  at  which  the
cyclotron  lines  begin to occur, and in this limit the cyclotron
quantum is of the order of or greater than the corresponding rest
energy. This is also  equivalent  to  the  requirement  that  the
de-Broglie  wavelength  is  of  the  order of or greater than the
Larmor radius of the particle in the magnetic field.

In  recent  years a lot of interesting work have been done on the
properties of dense  astrophysical  and  cosmological  matter  in
presence  of  strong  magnetic  field [15-17]. How the primordial
magnetic field affects the expansion rate  of  the  Universe  and
synthesis of light elements in the early Universe micro-second after
Big Bang has recently been
done  by  Schramm et al [18] (see also ref. [19]). In a series of
publication we have also reported the work done  by  us  on  the
effect  of  strong magnetic field on SQM using both MIT bag model
as well as D3QM model of confinement [20-23]. We have  also  done
some  work  on  the  magnetostatics of superconducting quark
stars and also on the stability and gross properties of strange
stars in presence of strong magnetic field [24].

The  aim  of the present paper is to investigate the stability and some
gross properties of quark matter in presence of  strong  magnetic
field. In section 2 we have studied the thermodynamic properties
of  bulk  SQM in presence of strong magnetic field. In section 3
the  thermal  nucleation  of  stable  quark matter
bubbles  in  compact  metastable  neutron  matter  at the core of
a  neutron  star  in  presence  of strong magnetic field has been
investigated. In section  4
the  properties  of co-existing bulk phases in presence of strong
magnetic field has been studied. Assuming the  possibility  of  a
metal-insulator   type   of   second   order  quark-hadron  phase
transition at the neutron star core at zero temperature,  we  have
investigated   the  effect  of  strong  magnetic  field  on  such
transition  in  section  5.  In  section  6,   the   paramagnetic
properties  of  SQM has been discussed in detail. Whereas, section 7 contains
conclusions of this work.

\medskip
\noindent {\bf{2. ~Bulk SQM in Strong Magnetic Fields}}
\medskip

To study the deconfining transition at the core and the stability
of  SQM  in presence of strong magnetic field, we have considered
the conventional MIT bag model. The limitation of this model has already
been discussed in the introduction.
For the sake of simplicity we are
assuming that quarks are moving freely within the system  and  as
usual the current masses of both $u$ and $d$-quarks are extremely
low  (in  our  actual  calculation we have taken current mass for
both of them to be $5$MeV, whereas  for  $s$-quark,  the  current
mass is taken to be $150$MeV).

For  a  constant  magnetic  field  along  the $z$-axis ($\vec B_m
=B_{m,(z)}=B_m=$ constant),  the  single  particle  energy  eigen
value is given by [25]
$$\epsilon_{k,n,s}^{(i)}=\left
[k^2+m_i^2+q_iB_m(2n+s+1)\right]^{1/2}\eqno(2.1)$$
where  $n=0,1,2,....,$  being  the  principal quantum numbers for
allowed Landau levels, $s=\pm1$ refers to spin up ($+$)  or  down
($-$)  states and $k$ is the component of particle momentum along
the direction of external magnetic field. Setting  $2\nu=2n+s+1$,
where  $\nu=0,1,2,...,$ we can rewrite the single particle energy
eigen value in the following form
$$\epsilon_\nu^{(i)}= \left [
k^2+m_i^2+q_iB_m2\nu\right]^{1/2}\eqno(2.2)$$
Now it is  very  easy  to  show  that  $\nu=0$  state  is  singly
degenerate  while  all  other  states  with $\nu\ne 0$ are doubly
degenerate.

The general expression for thermodynamic potential of the  system
is given by
$$\eqalignno{   &\Omega   =   -T   \ln   Z\cr  &=  \sum_i\left  [
\Omega_{i,V}(T,\mu_i)         V+\Omega_{i,S}(T,\mu_i)          S+
\Omega_{i,C}(T,\mu_i) C\right ]+BV &(2.3)}$$
where the sum is over $u$, $d$, $s$-quarks and electron ($e$).

The first term on the second line of the above  equation  is  the
volume contribution, whose explicit form is given by [26]
$$\Omega_{i,V}(T,\mu_i)=-{{Tg_i}\over{(2\pi)^3}}  \int  d^3k  \ln
\left ( 1+ \exp(\beta(\mu_i-\epsilon_i))\right )\eqno(2.4)$$
where $g_i$ is the degeneracy  of  the  $i$-th  species  ($=6$  for
quarks  and  $2$  for electron) and $V$ is the volume occupied by
the system.

The second term comes from the surface contribution and is  given
by [27]
$$\eqalignno{\Omega_{i,S}(T,\mu_i)={{g_iT}\over{64\pi^2}} \int
{{d^3k}\over{\mid \vec k\mid}}&\left [1- {{2}\over{\pi}} \tan^{-1}
\left ({{k}\over{m_i}}\right ) \right ]\cr
&\ln\left (1+\exp(\beta(\mu_i-\epsilon_i))\right )=\sigma&(2.5)}$$
and  $S$  is the area of the surface enclosing the volume $V$ and
$\sigma$ is the surface energy per unit area or the surface  tension
of quark matter.

The  last term corresponds to curvature correction, which becomes
zero for a plane interface separating two  phases.  The  explicit
form for curvature term is given by [28]
$$\eqalignno{\Omega_{i,C}(T,\mu_i)={{Tg_i}\over{48\pi^3}}    \int
{{d^3k}  \over{\mid\vec{k}\mid^2}}  &\ln  \left  [1+\exp   (\beta(\mu_i
-\epsilon_i))    \right    ]   \cr   &\left   [   1-{{3}\over{2}}
{{k}\over{m_i}}  \left  (   {{\pi}\over{2}}   -\tan^{-1}   \left
({{k}\over {m_i}} \right ) \right ) \right ] &(2.6)}$$
where  $C$  is  the length of the line element drawn on the surface
$S$.

At  finite  $T\sim\mu$,  one  has  to  consider  the  presence of
anti-quarks and positrons in  the  system.  We  assume  that  for
anti-particles   $\bar  \mu_i=-\mu_i$,  i.e.  they  are  also  in
chemical equilibrium with respect to annihilation processes.  Now
we  would like to see the necessary changes for these expressions
which has to be incorporated, if a strong external magnetic field
is present in  the system. With these modified form of thermodynamic
potentials, we shall investigate the magnetism arising from quantization
of  orbital  motion  of  charged  particles in presence of strong
magnetic field, known as Landau Diamagnetism. It is well known
that in presence of  a  magnetic
field  along $z$-axis, the path of the charged particle will be a
regular helix whose  axis  lies  along  the  $z$-axis  and  whose
projection  on  $x-y$ plane is a circle. If the magnetic field is
uniform, both the linear velocity along the field  direction  and
the  angular  velocity  in  the $x-y$ plane will be constant, the
latter arises from the constant Lorentz force experienced by  the
particle.  Quantum  mechanically  the  energy associated with the
circular motion in the $x-y$  plane  is  quantized  in  units  of
$2q_iB_m$.  the  energy  associated  with the linear motion along
$z$-axis is also quantized; but in view of the smallness  of  the
energy  intervals,  they may be taken as continuous variables. We
thus have eqn.(2.1) or (2.2)  as  single  particle  energy  eigen
value.  Now these magnetized energy levels are degenerate because
they result from an almost continuous set of zero  field  levels.
All   these   levels   for  which  the  values  of  the  quantity
$k_x^2+k_y^2$ lies between $2q_iB_m\nu$ and $2q_iB_m(\nu+1)$  now
coalesce  together  into  a  single  level  characterized  by the
quantum number $\nu$. The number of these levels is given by
$${{S}\over{(2\pi)^2}}  \int\int  dk_xdk_y={{Sq_iB_m}\over{2\pi}}
\eqno(2.7)$$
here  $S$  is  the  area  of  the  orbit in the $x-y$ plane. This
expression is independent of $\nu$. Then in the integral  of  the
form  $\int d^3k f(k)$, we can replace $\int\int dk_x dk_y$ by the
expression given above, whereas the limit of $k_z$,  which  is  a
continuous  variable, ranges from $-\infty$ to $+\infty$. Then we
can rewrite the eqns.(2.4) to (2.6) in presence  of  strong  magnetic
field  in the following form
$$\Omega_{i,V}=-T{{q_ig_iB_m}\over{2\pi^2}}   \sum_{\nu=0}^\infty
\int_0^\infty         dk_z         \ln\left         (1+\exp(\beta
(\mu_i-\epsilon_i^{(\nu)}))\right )\eqno(2.8),$$

$$\eqalignno{      \Omega_{i,S}=T      {{q_ig_iB_m}\over{16\pi}}
&\sum_{\nu=0}^\infty                                \int_0^\infty
{{dk_z}\over{\sqrt{(k_z^2+k_{\perp,(i)}^2)}}}   \ln\left  (1+\exp(\beta
(\mu_i-\epsilon_i^{(\nu)}))\right )\cr &\left  [1-{{2}\over{\pi}}
\tan^{-1} \left( {{k}\over{m_i}} \right ) \right ] &(2.9)}$$
and
$$\eqalignno{      \Omega_{i,C}=T      {{q_ig_iB_m}\over{12\pi^2}}
&\sum_{\nu=0}^\infty                                \int_0^\infty
{{dk_z}\over{(k_z^2+k_{\perp,(i)}^2)}}   \ln\left  (1+\exp(\beta
(\mu_i-\epsilon_i^{(\nu)}))\right  )\cr  &\left   [1-{{3}\over{2}}
{{k}\over{m_i}} \left ( {{\pi}\over{2}} -
\tan^{-1}  \left(  {{k}\over{m_i}}  \right  )\right  )  \right  ]
&(2.10)}$$
where $k_\perp^{(i)}=2\nu q_iB_m$

Since  the  surface  and  curvature  terms play significant roles
during quark bubble nucleation in dense neutron matter, and are not
at all important for a bulk quark matter system, we  shall
study  their  dependence  on  strong  magnetic  field in the next
section, when  we  shall  discuss  thermal  nucleation  of  quark
droplets in presence of strong magnetic field. In this section we
shall consider only the bulk term.

We  shall  use  eqn.(2.4)  for $s$ and $\bar s$ and eqn.(2.8) for
$u,d$  and  $e$  and  their   anti   particles.   In   eqn.(2.4),
$\epsilon_i=(k^2+m_i^2)^{1/2}$, whereas in eqn.(2.8), it is given
by eqn.(2.2).

Then from the well known thermodynamic relations [26], the
expression for kinetic pressure of the system is given by
$$P=-\sum_i \Omega_{i,V}\eqno(2.11),$$
the number density of the $i$th species is given by
$$n_i=-\left     (     {{\partial     \Omega_{i,V}}\over{\partial
\mu_i}}\right )_T\eqno(2.12)$$
and finally the expression for  the  corresponding  free   energy
density  is given by
$$U_i=\Omega_{i,V}+\mu_i n_i       -T\left       (      {{\partial
\Omega_{i,V}}\over{\partial T}} \right )_{\mu_i}\eqno(2.13)$$
The last term in eqn. (2.13) comes from the non-zero  entropy
of  the system, which is zero for $T=0$. Then the total energy
density of the confined SQM is given by
$$U=\sum_i U_i +B \eqno(2.14)$$

To  evaluate   all   these   thermodynamic  quantities as defined
above  for  a  given
temperature  and  bag  parameter, we need chemical potentials for
all the constituents   present   in  the  system.  Assuming  the
condition  of   $\beta$-equilibrium, we have
$$\mu_d=\mu_s=\mu ~~~~(\rm{say})~~~~~\eqno(2.15)$$
and
$$\mu_u=\mu-\mu_e\eqno(2.16)$$
Whereas the  charge neutrality condition gives
$$2n_u-n_d-n_s-3n_e=0\eqno(2.17)$$

Again the baryon number density of the system is given by
$$n_B={{1}\over{3}}(n_u+n_d+n_s)\eqno(2.18)$$
which is considered as a constant parameter.

We  have solved  these  equations numerically to obtain chemical
potentials for all the specieses  for  various  values  of  $B_m$
(including  zero)  and then evaluate the thermodynamic quantities
of the SQM system.

Since for any non-zero temperature $T>>0$, analytical expressions
for  the  thermodynamic  quantities can not be obtained, we shall
give zero temperature  and  for  the  sake  of  illustration  low
temperature analytical expressions. Whereas in actual calculation
for  $T\ne  0$,  we  shall use numerical solutions of the general
expressions.

Now for $T=0$, we have the number density of  the  component  $i$
($i=u,d$ or $e$)
$$n_i=n_i^*\left                                                [
\zeta_i^{1/2}+2\sum_{\nu=1}^{[\nu_{max}^{(i)}]}(\zeta_i-\nu)^{1/2}\right
]\eqno(2.19)$$
where
$$n_i^*={{q_iB_m}\over{2^{1/2}\pi^2}} g_i\eqno(2.20)$$
$$\zeta_i={{\mu_i^2-m_i^2}\over{2q_iB_m}}\eqno(2.21)$$
and  $[\nu_{max}^{(i)}]$  is  the  greatest  integer not exceeding
$\zeta_i$. The $\nu$ limit becomes finite  for  $T\rightarrow  0$
limit  only,  when  the  maximum  available  energy  of a particle
$\approx$ its Fermi energy.
On the other hand,  for $s$-quark, the expression for number density is given
by the usual expression
$$n_s={{(\mu_s^2-m_s^2)^{3/2}}\over{\pi^2}}\eqno(2.22)$$

In  fig.(1)  we  have shown the variation of number densities for
$u$,  $d$,  $s$-quarks  and  electron  with  the  magnetic  field
intensity  $B_m$.  At low enough magnetic field strength, all the
number densities remain almost constant and are  equal  to  their
zero   field   values.  For  relatively  higher  field  strengths
$>10^3\times  B_m^{(c)(e)}$G,  electron  density  increases  with
$B_m$.  The  $s$-quark density decreases with $B_m$ at relatively
higher values than that of electron and becomes negligibly  small
for  extremely  large field strength $>10^5\times B_m^{(c)(e)}$G.
On the other hand, $u$ quark density becomes almost two times the
zero field value at this extremely large field strength.  Whereas
the  $d$-quark  density remains almost constant within this range
of magnetic field. But all the number densities show  oscillating
behavior  with  the  increase  of  magnetic  field intensity. The
number densities show  oscillating  nature,  as  the  consecutive
Landau   levels  for  different  constituents  pass  through  the
corresponding Fermi levels. The  electron  density  increases  to
$\sim  10^4$  times  zero field value at extremely large magnetic
field. One can very  easily  check  that  these  oscillating  and
increasing  and  /  or decreasing natures of number densities are
consistent with the charge neutrality  and  constancy  of  baryon
number  density  conditions.  The  physical  meaning of change in
$s$-quark  density  and  electronic  density  is  the  shift   of
$\beta$-equilibrium  point  in presence of strong magnetic field.
As a cross check one should try to see how do the rates  of  weak
processes  in  which  $s$-quark  and / or electron are created or
annihilated, change in  presence  of  strong  magnetic  field  (a
detail  analysis  on  the effect of strong magnetic field on weak
and electromagnetic processes in SQM will be reported somewhere else [29])

The  expression  for kinetic pressure and free energy density for
the $i$th species are given by ($i=u,d$ and $e$)
\vfil\eject
$$\eqalignno{&P_i=-\Omega_{i,V}\cr &={{q_ig_iB_m}\over{2\pi^2}}
\sum_{\nu=0}^{[\nu_{max}^{(i)}]}
\big  [ {{1}\over{2}}\mu_i(\mu_i^2-M_\nu^{(i)2})^{1/2}\cr &-
{{1}\over{2}} M_\nu^{(i)2}  \ln \left \{ {{\mu_i+(\mu_i^2-M_\nu^{(i)2})^{1/2}
}\over {M_\nu^{(i)} }}\right \}\big ]&(2.23)}$$
and
$$\eqalignno{&\epsilon_i =\Omega_{i,V}+\mu_i n_i\cr
&={{q_ig_iB_m}\over{2\pi^2}} \sum_{\nu=0}^{[\nu_{max}]}
\Biggl  [{{1}\over{2}}\mu_i(\mu_i^2-M_\nu^{(i)2})^{1/2}\cr &+
{{1}\over{2}} M_\nu^{(i)2}  \ln \left \{ {{\mu_i+(\mu_i^2-M_\nu^{(i)2})^{1/2}
}\over {M_\nu^{(i)} }}\right \}\Biggr ]&(2.24)}$$
where  $M_\nu^{(i)}=(m_i^2+2\nu  q_i  B_m)^{1/2}$.  Whereas  for
$s$-quark, they are given by the usual expressions
$$\eqalignno{P_s ={{1}\over{8\pi^2}} \big [ 2\mu_s
(\mu_s^2-m_s^2)^{3/2} &-3m_s^2\mu_s (\mu_s^2-m_s^2)^{1/2}\cr &+
3m_s^4 \ln \left \{ {{\mu_s+(\mu_s^2-m_s^2)^{1/2}
}\over {m_s} }\right \}\big ] &(2.25)}$$
and
$$\eqalignno{\epsilon_s=
{{3}\over{8\pi^2}} \big  [ 2\mu_s^3
(\mu_s^2-m_s^2)^{1/2} &-m_s^2\mu_s (\mu_s^2-m_s^2)^{1/2} \cr &-
m_s^4 \ln \left \{ {{\mu_s+(\mu_s^2-m_s^2)^{1/2}
}\over {m_s} }\right \}\big ]&(2.26)}$$
These two expressions are without the bag pressure or vacuum  energy
density.  Here  we  have also not taken magnetic pressure and magnetic
energy density  into  account.  They  are   given   by the  usual
expressions,
$\pm  B_m^2/8\pi$,  where $+$ sign is for magnetic energy density
and $-$ sign is for magnetic pressure.
Now  for $B_m=10^{15}$G, the magnetic energy density
$\epsilon_{15}=32.283$MeV$^4$, which is too  small compared  with
the free energy density $\sim 10^6$ MeV$^4$,  whereas  for
$B_m\sim 10^{18}$G,  it  is given by $\epsilon_{18}=10^6\times \epsilon_{15}$.
The  same  conclusion  is  also valid for magnetic pressure of the
system.
Therefore, magnetic energy density as well as the magnetic pressure
will  be  comparable  with  the
particle   free   energy   density   and  the  kinetic   pressure
respectively if  and only if the magnetic
field intensity is too high. For $B_m\geq 10^{20}$G, the magnetic
energy density as well as the magnetic pressure play the main
roles. This magnetic field strength is however, too high to realize
at  the  core  of neutron star or any other stellar objects of our
interest.

To study the stability of bulk SQM in presence of strong magnetic
field we shall use the stability condition $\sum_iP_i-B=0$, where
$i=u,d,s$ and  $e$,  and  solve  the  eqns.(2.15)-(2.18)
numerically  for  $T=0$.  Since  $\nu_{max}^{(i)}$  is  a function of
$\mu_i$, we have solved these  equations  self  consistently  for
$B_m\neq  0$. In fig.(2) we have plotted energy per baryon of SQM
against the  baryon  number  density.  The  upper  curve  is  for
$B_m=0$,   whereas   the   lower   one   is  for  $B_m=10^3\times
B_m^{(c)(e)}$. This figure shows clearly  that  the  presence  of
strong magnetic field makes SQM energetically more stable.

We  shall  now  write  down  for the sake of completeness the low
temperature analytical expressions for  thermodynamic  parameters.
The number density is given by
$$n_i=n_i^{(0)} -{{q_ig_iB_m}\over{12}} T^2
\sum_{\nu=0}^{[\nu_{max}^{(i)}]} (\mu_i^2-M_\nu^{(i)2} )^{1/2}
{{M_\nu^{(i)2}}\over{(\mu_i^2-M_\nu^{(i)2})^2}}\eqno(2.27)$$
for $i=u,d$ and $e$, and as before, for $s$-quark it is given by
$$n_s=n_s^{(0)}
+{{T^2}\over{6}}{{2\mu_s^2-m_s^2}\over{(\mu_s^2-m_s^2)^{1/2}}}
\eqno(2.28)$$
The expressions for pressure for $u, d$ and $e$ is given  by  the
general expression
$$P_i=P_i^{(0)}+{{q_ig_iB_m}\over{12}} T^2
\sum_{\nu=0}^{[\nu_{max}^{(i)}]}
{{\mu_i}\over{(\mu_i^2-M_\nu^{(i)2})^2}}\eqno(2.29)$$
and for $s$-quark it is given by
$$P_s=P_s^{(0)}+{{T^2}\over{2}}\mu_s(\mu_s^2-m_s^2)^{1/2}\eqno(2.30)$$
In  all these eqns., the supescripts $"0"$ refer
to zero temperature cases. In this particular low temperature
case the free energy density for the $i$th species is given by
$$\epsilon_i=-P_i+\mu_in_i +Ts_i\eqno(2.31)$$
where  $i=u,d,s$ or $e$.
here  the last term comes from non-zero entropy of the system and
is given by
$$Ts_i=-T\left ( {{\partial \Omega_{i,V}}\over{\partial  T}}\right
)_{\mu_i}={{q_ig_iB_m}\over{6}}T^2
\sum_{\nu=0}^{[\nu_{max}^{(i)}]}
{{\mu_i}\over{(\mu_i^2-M_\nu^{(i)2})^{1/2}}}\eqno(2.32)$$
for $i=u,d$ and $e$, whereas for $s$-quark it is given by
$$Ts_s=-T\left ( {{\partial \Omega_{s,V}}\over{\partial  T}}\right
)_{\mu_s}=T^2\mu_s(\mu_s^2-m_s^2)^{1/2}\eqno(2.33)$$

In  figs.(3) and (4) we have shown the equation of states for SQM
at $T=0$ and $40$MeV respectively. In both these figs. the dotted
curves   are   for  $B_m=0$,  whereas  the  solid  ones  are  for
$B_m=10^3\times B_m^{(c)(e)}$. For the sake of comparison we have
plotted  B-J equation of state [14] in fig.(3) and is indicated by the
symbol $bj$. These two figs. show that in presence of strong magnetic
field the equation of state of SQM changes significantly. In both
these figs. we have taken $B^{1/4}=0$, which is a constant scaling
factor.  Fig.(5) shows the variation
of  energy  per  baryon  of  stable  SQM  with the magnetic field
strength for $n_B=2.5n_0$  and  $T=0$.  This  fig.   shows   that
for low $B_m$, the
energy  per  baryon  remains almost constant and
rises  sharply as $B_m$ exceeds $ 10^4
\times  B_m^{(c)(e)}$,  and  the  matter   becomes   energetically
unstable.  In  fig.(6)  we  have plotted the variation of kinetic
pressure of SQM with the magnetic field intensity for $B^{1/4}=0$
and  for the same  value  of  $n_B$  as  before.  This fig. shows
that for relatively low values of  magnetic  field  strength
the  kinetic  pressure
initially
remains almost independent of magnetic field strength
and then decreases sharply as $B_m$ exceeds
$  10^4  \times   B_m^{(c)(e)}$  and  finally becomes negative for
extremely large values for $B_m$. Therefore for  extremely  large
magnetic field strength, the system becomes energetically as well
as mechanically unstable (these two conclusions are valid if the magnetic field is
is confined within the quark matter region only, which is possibly not true).

\medskip
\noindent {\bf{3. ~Nucleation of  Quark  Bubble  in  Compact  Neutron
matter}}
\medskip

In this section we shall investigate the first order quark-hadron
phase  transition initiated by the nucleation of droplets of quark matter
in  presence  of  strong  magnetic  field.  In
particular  we shall study the effect of strong magnetic field on
the nucleation of quark droplets in dense neutron  matter.  Since
the  surface  and  curvature  energies  of  the quark bubble play
crucial role in droplet nucleation we would like to see how  these
two  quantities  are  modified by the presence of strong magnetic
field.

Now the  nucleation rate of stable quark matter bubble  in
metastable  neutron  matter per unit volume due to thermal fluctuation is
given by [30]
$$I=I_0   \exp   \left  (-W_m/T\right  )\approx  T^4  \exp  \left
(-W_m/T\right )\eqno(3.1)$$
where  $W_m$  is  the  minimum  thermodynamic  work to be done to
create a critical quark droplet and is given by [31]
$$W_m={{4}\over{3}}  \pi  {{\sigma^3}\over{(\Delta  P)^2}}   [2+2
(1+b)^{3/2} +3b]\eqno(3.2)$$
where  $\sigma=\sigma_q+\sigma_n$ the surface tension and $\Delta
P=P_q-P_n$ is the pressure difference. Here $\Delta P=P_q-P_n$ is
a positive quantity and both $\sigma_q$ and $\sigma_n$  are  also
positive,      $b=2\gamma      (\Delta      P)/\sigma^2$,     and
$\gamma=\gamma_q-\gamma_n$,    stands   for   curvature    energy
density.
Here  phase "$q$" being a droplet of quark matter and phase
"$n$" stands for the metastable neutron matter. In  eqn.(3.1),  the
pre-exponential  factor  is  chosen to be $T^4$, where $T$ is the
temperature of the metastable medium (for an exact expression see
ref. [32]). In eqn.(3.2) we have assumed that the two phases are  in
chemical   equilibrium   ($\Delta   \mu=0$).   Since  the  bubble
nucleation  time   $\tau_{\rm{bubble}}\approx   10^{-23}~~   {\rm
{sec}}~~\approx\tau_{\rm{strong}}$,  the  strong interaction time
scale, the creation of  strange  quarks  through  weak  processes
within  the quark droplets may be ignored. Therefore the constituents
of quark droplets are  only $u$ and $d$ quarks. Since the  temperature
($\sim  4-5$ MeV) $<<$ quark chemical potential ($\sim 300$ MeV),
the presence of anti-quarks can also be  ignored.  On  the  other
hand,  if  we  assume  the  presence  of  hyperons at the core of
neutron   star   (which   is   believed   to   be   true),   then
immediately after the
phase transition to quark matter, $s$ quarks will
also be present in the quark bubble.  However,  the  quantum  mechanical
effect  of strong magnetic on $s$ quark part of strangelet can be
ignored as long as the magnetic field strength is $\leq  10^{20}$
G, which has already been discussed.
Therefore  in  the  surface  energy  term  of strangelet a finite
contribution will come from $s$ quark part and is given  by
eqn.(2.5).  On  the other hand the surface tension or the surface
energy per unit  area  for  $u$  and  $d$  quarks  are  given  by
eqn.(2.9).  It  is obvious from this eqn. that the surface energy
diverges  logarithmically  in  the  infra red  limit
(i.e., $k_z\rightarrow 0$)   for $\nu=0$, i.e. for the ground state Landau
level.  To show  this more
explicitly, we shall make zero temperature approximation. In this
case
the  upper  limit  of  the  $\nu$-sum  can  be  obtained from the
condition
$$k_z^2=\mu_i^2-m_i^2-2\nu q_i B_m \geq 0\eqno(3.3)$$
This gives
$$\nu      \leq   {{\mu_i^2-m_i^2}\over{2q_iB_m}}~~
= [\nu_{max}^{(i)}] ~~(\rm{say})\eqno(3.4)$$
which    is     the     largest     integer     not     exceeding
$(\mu_i^2-m_i^2)/(2q_i B_m)$.
Therefore  as has been mentioned earlier the  upper  limit
$[\nu_{\rm{max}}^{(i)}]$
can not be same for $u$ and $d$ quarks.
Now  to  visualize  the  effect  of  magnetic  field  on  the
nucleation   rate  of  quark  droplet,  we  shall  first  rewrite
eqn.(2.9) in the form
$$\sigma={{TB_m}\over{16\pi}}           \sum_{i=u,d}          g_i
q_i\sum_{\nu=0}^\infty                              \int_0^\infty
{{dk_z}\over{\sqrt{k_z^2+k_{\perp     (i)}^2}}}     \ln     \left
[1+\exp\left  (-{{\epsilon_i^{(\nu)}-\mu_i}\over{T}}   \right   )
\right ] G \eqno(3.5)$$
where
$$G=             \left             [1-            {{2}\over{\pi}}
\tan^{-1}\left({{k}\over{m_i}}\right ) \right ]$$
for MIT bag model and is equal to unity for D3QM model.
Now we shall evaluate integral (3.5) by
parts  for  $G=1$  and take $T\rightarrow 0$ limit, then
$$\eqalignno  {\sigma(G=1)&={{B_m}\over{16\pi}}  \sum_{i=u,d} g_i
q_i \big \{ \sum_{\nu=0}^{\nu_{\rm{max}}^{(i)}} \int_0^{k_{F_i}}  {{k_z
dk_z}\over  {\sqrt{k_z^2+m_i^2+k_{\perp,(i)}^2}}}  \ln(k_z+\sqrt{
k_z^2+k_{\perp,(i)}^2  })\cr   &-   \sum_{\nu=0}^{\nu_{\rm{max}}}
\ln(k_{\perp,(i)})  (\mu_i-\sqrt{k_{\perp,(i)}^2+m_i^2})\big  \}
 &(3.6)}$$
It   is   easy   to   check  that  eqn.(3.6)  is  also  a  part  of
$\sigma$, given by eqn.(3.5)  for   which   $G\neq   1$.   Since   $\mu_i>
\sqrt{k_{\perp,  (i)}^2+m_i^2}$, therefore, for $\nu=0$ the second
term of eqn.(3.6) becomes $+\infty$. The other part  of  integral
(3.6)  has  been  evaluated  numerically and found to be a finite
number. Similarly the integral (3.5) with the  second  part  of
$G$  (which  is  $\ne  1$) has also been obtained numerically and
is found to be  finite. Therefore the diverging property  of
$\sigma(G=1)$  for  $\nu=0$  is  also  true for the whole surface
energy / area, given by eqn.(3.5).
The term $\nu=0$ corresponds to the lowest Landau level.  If  the
magnetic  field  is too strong only this level will be populated,
which is also obvious from eqn.(3.4).
Although  we  have  assumed here $T\rightarrow 0$, this important
conclusion is equally valid for any finite $T$. Since both  $P_n$
and  $P_q$  are  finite even in presence of strong magnetic field
and $T$ is  also  finite,  we  have  from  eqn.(3.1)  $I=  0,$
which means  the thermal nucleation rate of droplet formation becomes
zero, i.e., there
can not be a  single  quark  droplet  formation  from  metastable
neutron matter due to  thermal fluctuation
at the  core,  if  the
magnetic   field   is  of  the  order  of  or  greater  than  the
corresponding  critical  value.  Therefore the formation of quark
droplet at the magnetized neutron star core  will be  controlled  by
some  other mechanism, namely, the quantum effects, triggering by
strangelet capture etc.

Following ref. [28], we shall now investigate the effect of  strong
magnetic  field  on  the quark curvature term which also plays an
equally  important  role  as  surface  tension   during   thermal
nucleation. As before, we can rewrite eqn.(2.10) in the form
$$\gamma =    {{T}\over{12\pi^2}}       \sum_{i=u,d} g_i      q_i
\sum_{\nu=0}^\infty                \int_0^\infty
{{dk_z}\over{k_z^2+k_{\perp,(i)}^2}}
\ln\left (1+\exp(-\beta(\epsilon_i-\mu_i))\right ) G\eqno(3.7)$$
where
$$G=1-{{3}\over{2}}        {{k}\over{m_i}}         ({{\pi}\over{2}}
-\tan^{-1}({{k}\over {m_i}} ))$$
for MIT bag model and equal to unity for D3QM model.

Now we shall evaluate eqn.(3.7) term by term. Consider
$$I_1=\int_0^\infty {{dk_z}\over{k_z^2+k_{\perp,(i)}^2}}
\ln\left (1+\exp(-\beta(\epsilon_i-\mu_i))\right )\eqno(3.8)$$
Obviously, this integral diverges in the infra red limit for $\nu=0$. Here
the divergence is $1/k_z$ type. To see it more explicitly, let us integrate
eqn.(3.8) by parts, then we have
$$I_1={{1}\over{Tk_{\perp,(i)}}}                   \int_0^\infty
\tan^{-1}({{k_z}\over{k_{\perp,(i)}}})         {{k_z         dk_z}\over
{(k_z^2+k_{\perp,(i)}^2 +m_i^2)^{1/2}}}
{{1}\over{\exp(\beta(\epsilon_i-\mu_i))+1}} \eqno(3.9)$$
which diverges for $\nu=0$, but the divergence is not logarithmic
($I_1\sim 1/\nu$ as $\nu \rightarrow 0$).

Next consider the second term
$$I_2=-{{3\pi}\over{4m_i}}        \int_0^\infty        {{dk_z}\over
{(k_z^2+k_{\perp,(i)}^2)^{1/2}}}
\ln\left (1+\exp(-\beta(\epsilon_i-\mu_i))\right )\eqno(3.10)$$
Which has logarithmic divergence in the infra red limit for $\nu=0$.
Integrating by parts, we have
$$\eqalignno{I_2 &={{3\pi}\over{4m_i}}[ \{ \ln(k_{\perp,(i)})
\ln\left (1+\exp(-\beta(\epsilon_i-\mu_i))\right )\}\cr
&- {{1}\over{T}} \int_0^\infty \ln(k_z+(k_z^2+k_{\perp,(i)}^2)^{1/2})
{{k_zdk_z}\over{(k_z^2+k_{\perp,(i)}^2+m_i^2)^{1/2}}} \cr
& \left [{{1}\over{\exp(\beta(\epsilon_i-\mu_i))+1}}\right ]]
&(3.11)}$$
For $\nu=0$, the first term diverges logarithmically, but  unlike
the  second  term of eqn.(3.6), it becomes $-\infty$. The  second
term of eqn.(3.11) and the last  term of eqn.(3.7)
have  been  checked  numerically and they remain  finite  for all
values of  $\nu$ in the infra red as well as ultra violet limits.
Now the divergences of first two terms of this eqn.
can not cancel each other.
The first divergence is much faster as  $\nu\rightarrow  0$  than
the second one, therefore the  overall  divergence  of  curvature
term (3.7) remains positive
as $\nu\rightarrow 0$. Now from eqn.(3.2) it  is  obvious  that  if
some  how  the quantity $b$ becomes finite (even zero), still the
effect  of  $\sigma$ makes $W_m$ infinitely  large.  Therefore, in
general, both the surface tension as well as curvature terms of a
quark matter bubble diverge in presence of strong magnetic field.
Physically it means, that an infinite amount of work  has  to  be
done  to  create  a  critical  quark  bubble.  As a consequence,
thermal nucleation of such bubble in  metastable  neutron  matter
will be completely forbidden.
It can only occur by some non-thermal mechanisms.
Therefore,  in  future, if we get some direct or indirect
experimental evidences for quark core and if the above conclusion
is assumed to be correct, then either the magnetic field of  such
objects  will  be  extremely low ($< B_m^{(c)}$) from its time of
birth, if quark matter is  assumed  to  be  produced  immediately
after  supernova  explosion,  or  it must have been produced much
latter, when the neutron star magnetic field  has  decayed  to  a
value  $B_m  <  B_m^{(c)}$  or  the  quark  droplets are produced
through some non-thermal means. Since all these divergences are appearing
in the infra red limit for $\nu=0$, one can in principle remove them by
introducing a lower cutoff for $k_z$. Ofcourse, in that case one has to think
the physical meaning of this lower cutoff of longitudinal momentum.

\medskip
\noindent {\bf{4. ~Co-existing Bulk Phases}}
\medskip

From the discussion of previous section, we have seen that it  is
impossible  to  have  a  mixed  phase of dense neutron matter and
droplets of quark matter in presence  of  strong  magnetic  field
of strength
greater than the corresponding critical value. In this section we
would like to see the effect of  strong  magnetic  field  on  the
coexisting bulk phases.

Assuming for the sake of simplicity that only neutrons present in
the hadronic phase and obey the free  hadronic  gas  equation  of
state;  then  the  conditions  to be satisfied by the co-existing
phases are
$$\mu_n=\mu_u+2\mu_d\eqno(4.1)$$
$$P_n=P_q\eqno(4.2)$$
and $$T_n=T_q=T_c\eqno(4.3)$$
We also assume that the condition of $\beta$-equilibrium is satisfied
in bulk quark matter phase.

In fig.(7) we have shown the phase diagram ($n_B$ vs $T$ diagram)
for such a mixed phase.
The dotted curve is for  $B_m=0$,  whereas  solid  curve  is  for
$B_m=10^3\times  B_m^{(c)(e)}$.  This figure shows that the phase
diagram changes significantly  in  presence  of  strong  magnetic
field.

This observation is  specially  important during  quark-hadron
phase  transition in the early Universe (where $n_B\approx 0$) if
it would have taken place in presence of  strong  magnetic  field
(!!) of our interest (for a detail discussion, see the in press paper
of ref. [23], where the effect of strong magnetic field on the baryon number
inhomogeneity during cosmic QCD phase transition has also been discussed).

\medskip
\noindent{\bf{5 ~Metal-Insulater Type of Transition}}
\medskip

Since the order of quark-hadron trasition is still not known exactly,
in  this  section  we  shall  consider  a metal insulator type of
second order quark-hadron phase transition at zero temperature in
presence  of strong magnetic field. Now as the density of neutron
matter at the core of  neutron  star  increases,  the  separation
between  two  neutrons decreases and at some critical value, when
the separation  becomes  $\sim  0.4$fm,  the  hard  core  radius,
neutrons  virtually  loose  their  individuality  and  quarks  can
percolate out, give rise to a color  metal.  This  transition  is
more  or  less  identical  with  the  metal  insulator transition
observed in condensed matter physics, which takes place  at  high
pressure.   Here   neutrons,  which  are  colorless objects play
the role of
neutral atoms of the insulator  placed  at  the
lattice points and orbital bound electrons may be  compared  with
the confined quarks.

Unlike  a  first  order  transition,  where the two phases are in
equilibrium at the critical point, here they  are  indistinguishable.
The  conditions  which  must  be  satisfied  in  this case at the
critical point are
$$\epsilon_n= \epsilon_q\eqno(5.1)$$
$$\mu_n=2\mu_d+\mu_u\eqno(5.2)$$
and of course
$$P_n=P_q\eqno(5.3)$$

Here we have assumed the presence of $u$ and $d$ quarks only,  the  numerical
solution  of  the eqns.(5.1-5.3) give critical baryon number density
$n_c=1.49n_0$ and $B^{1/4}=146.52$MeV for $B_m=0$. But in presence of
strong magnetic field, both the critical baryon number density and the
corresponding bag pressure depend on the strength of magnetic field.
In fig.(8)  we
have  shown  the  variation  of  $n_c$  with  the  magnetic field
intensity $B_m$ and in fig.(9) the  same  type  of  variation  is
shown for $B^{1/4}$.
These figs. show that  for  low  magnetic  field  strength,  both  the
critical   density   and  bag  pressure are almost independent of
$B_m$ and
very close to their  zero magnetic field values. But  for  large
magnetic   field  strength  $> 10^4\times B_m^{(c)(e)}$G,
both  these  quantities  increase
sharply. These figs. also show  that for extremely large  magnetic
field  strength, the critical density becomes too high to achieve
at the core of neutron star. Which implies that for an extremely
strong  magnetic field at the neutron star core, both the first
order transition through the
nucleation of quark bubbles and a second order metal insulator type
of  quark  hadron  transition  are impossible.

\noindent {\bf{ 6. ~Pauli Paramagnetism of Quark Matter}}
\medskip

As has been mentioned in the introduction, that
the magnetism of quark gas at $T=0$ consists of  two  parts:  a
paramagnetic  part  due  to  the  intrinsic spin magnetic moment, known as
Pauli  paramagnetism  and  a  diamagnetic  part   called   Landau
diamagnetism,  arises from the quantization of the orbital motion
of the quarks in the strong magnetic field. The second  part  has
already  been  discussed.  In this section we shall consider
the Pauli paramagnetism of quark matter.

It is known that the quarks have non-zero magnetic dipole moment,
given by
$\eta_u=1.852\eta_N$,          $\eta_d=-0.972\eta_N$          and
$\eta_s=-0.581\eta_N$,      where      $\eta_N=e/2m_p=3.152\times
10^{-14}$MeV/Tesla,  the  nuclear magneton. Whereas for electron,
it     is     given     by     $\eta_e=1.00116\eta_B$,      where
$\eta_B=e/2m_e=5.788\times 10^{-11}$MeV/Tesla, the Bohr magneton.

Therefore  in presence of a magnetic field of strength $B_m$, the
magnetic potential energy of the $i$th component is $-\eta_iB_m$.
Where $i$ stands for $u$, $d$ or $s$-quarks or electron.
Then the corresponding number density is given by [33]
$$n_i={{g_i}\over{2\pi^2}}                   \left                  [
\int_{\epsilon=m_i}^{\epsilon_i(B_m)} k^2dk +
\int_{\epsilon=m_i+2\eta_iB_m}^{\epsilon_i(B_m)}   k^2dk   \right
]\eqno(6.1)$$
which gives after some simple algebraic manipulation
$$n_i={{2\eta_i     B_m     (2\eta_iB_m+2m_i)}\over     {6\pi^2}}
(2x_i^{3/2}-1) \eqno(6.2)$$
where
$$x_i={{\epsilon_i^2  (B_m)- m_i^2} \over {2 \eta_i B_m (2\eta_i
B_m + 2m_i)}}\eqno(6.3)$$
and $\epsilon_i$ is the single particle energy. From  eqn.(6.3)  one
can write down
$$\epsilon_i^2(B_m)=m_i^2+x_i[R_{3/2}]^{-2/3}
(\epsilon_{i,0}^2-m_i^2)\eqno(6.4)$$
where $\epsilon_{i,0}$ is the single particle energy  in  absence
of magnetic field, given by
$$\epsilon_{i,0}=m_i^2+(3\pi^2 n_i)^{2/3}\eqno(6.5)$$
and
$$R_{3/2}(x_i)={{1}\over{2}}(2x^{3/2}-1)\eqno(6.6)$$
Now the total particle number density for the $i$th component can
be decompossed into two parts; with the magnetic moment along  the
direction of magnetic field and opposite to it, then we have
$$n_i=n_i^{(\uparrow)}(B_m)+n_i^{(\downarrow)}(B_m)\eqno(6.7)$$
where
$$n_i^{(\uparrow)}(B_m)={{n_i}\over{2}}
x_i^{3/2}[R_{3/2}(x_i)]^{-1}\eqno(6.8)$$
the  number density for the $i$th component with the direction of
magnetic dipole moment along the direction of  external  magnetic
field and
$$n_i^{(\downarrow)}(B_m)={{n_i}\over{2}}
x_i^{3/2}[R_{3/2}(x_i)]^{-1}(1-x_i^{-3/2})\eqno(6.9)$$
the  same  number density as above, but the direction of magnetic
dipole moment is now
opposite to the direction of external field.
In absence of magnetic field,  or  for  $B_m\rightarrow  0$,  the
quantity $x_i=\infty$, which gives
$$\epsilon_i=\epsilon_{i,0},
~~~~~~n_i^{(\uparrow)}=n_i^{(\downarrow)}\eqno(6.10)$$
The second relation implies equal occupation probability  for  up  and  down
spin states. On the other hand for $x_i=1$, we have
$$n_i^{(\uparrow)}=n_i,~~~                          {\rm{and}}~~~
n_i^{(\downarrow)}=0\eqno(6.11),$$
which   is  the  complete  saturation  of the  spin  states along
the direction of magnetic field.   In
this condition
$$\epsilon_i^2=m_i^2+2^{2/3}(\epsilon_{i,0}^2-m_i^2)\eqno(6.12)$$
The   threshold  magnetic  field strength for complete saturation
of   $i$th  component  can  be  obtained  by  putting  $x_i=1$ in
eqn.(6.3),
which gives
$$B_m^{(th)(i)}={{\left               [              2^{2/3}\left
({{6\pi^2 n_i}\over{g_i}}\right   )^{2/3}   +m_i^2\right    ]^{1/2}
-m_i}\over {2\eta_i}}\eqno(6.13)$$
For  $n_i=2.5n_B$, the threshold values for $u, d$ and $s$ quarks
are $3.4\times 10^{19}$G, $6.46\times 10^{19}$G  and  $7.59\times
10^{19}$G, respectively, which are of course not too low.

The  energy density for such a Pauli paramagnetic system is given
by
$$U=\sum_i U_i\eqno(6.14)$$
where
$$U_i(B_m)={{g_i}\over{2\pi^2}}                   \left                  [
\int_{\epsilon=m_i}^{\epsilon_i(B_m)} \epsilon k^2dk +
\int_{\epsilon=m_i+2\eta_iB_m}^{\epsilon_i(B_m)}\epsilon  k^2dk \right ]
\eqno(6.15)$$
which after some straight forward  algebraic manipulation becomes
$$\eqalignno{U_i(B_m)     &={{g_i}\over{16\pi^2}}     \big     [
2(\epsilon_i^2-m_i^2)^{1/2}    (2\epsilon_i^2-m_i^2)   \epsilon_i
-2m_i^4 \ln \left ( {{\epsilon_i+ (\epsilon_i^2-m_i^2)^{1/2}}\over
{m_i}}\right ) \cr &-  (y_i(y_i+2m_i))  (m_i^3+5  m_i^2y_i  +6m_i
y_i^2
+2y_i^3)\cr & +m_i^4 \ln \left ( 1+{{y_i +(y_i (2m_i+y_i))^{1/2}} \over {
m_i}} \right ) \big ] &(6.16)}$$
where $y_i=2\eta_i B_m$

The kinetic pressure is given by
$$P=\sum_i P_i\eqno(6.17)$$
where
$$P_i(B_m)={{g_i}\over{6\pi^2}}\left [ \int_{\epsilon=m_i}^{\epsilon(B_m)}
{{k^4dk}\over{(k^2+m_i^2)^{1/2}}}+
\int_{\epsilon=m_i+2\eta_i}^{\epsilon(B_m)}
{{k^4dk}\over{(k^2+m_i^2)^{1/2}}}\right ] \eqno(6.18)$$
after some algbraic simplification it reduces to
$$\eqalignno{P_i(B_m)     &={{g_i}\over{48\pi^2}}     \big     [
2(\epsilon_i^2-m_i^2)^{1/2}    (2\epsilon_i^2-m_i^2)   \epsilon_i
+6m_i^4 \ln \left ( {{\epsilon_i+ (\epsilon_i^2-m_i^2)^{1/2}}\over
{m_i}}\right ) \cr &- (y_i(y_i+2m_i)) (-3m_i^3+ m_i^2y_i +6m_i y_i^2
+2y_i^3)\cr & -3m_i^4 \ln \left ( 1+{{y_i +(y_i (2m_i+y_i))^{1/2}} \over {
m_i}} \right ) \big ]&(6.19)} $$
In  the  field  free  case  ($B_m=0$),  $y_i=0$,  then  both  the
expressions  for  energy  density  and  pressure  reduce  to  the
corresponding zero field
expressions, with $\epsilon_i$ replace by $\epsilon_i^{(0)}$.
On the other hand    for    the    saturation    condition,    $x_i=1$ and
$y_i=\epsilon_i-m_i$, then we have
$$U_i(x_i=1)    ={{g_i}\over{16\pi^2}}     \big     [
2(\epsilon_i^2-m_i^2)^{1/2}    (2\epsilon_i^2-m_i^2)   \epsilon_i
-2m_i^4 \ln \left ( {{\epsilon_i +(\epsilon_i^2-m_i^2)^{1/2}}\over
{m_i}}\right ) \big ]\eqno(6.20) $$
where   in   this   expression   $\epsilon_i$   is    given    by
eqn.(6.12).
Similarly the expression for pressure is given by
$$P_i(x_i=1)    ={{g_i}\over{48\pi^2}}     \big     [
2(\epsilon_i^2-m_i^2)^{1/2}    (2\epsilon_i^2-m_i^2)   \epsilon_i
+6m_i^4 \ln \left ( {{\epsilon_i+ (\epsilon_i^2-m_i^2)^{1/2}}\over
{m_i}}\right ) \big ] \eqno(6.21)$$

Now the bulk interaction energy per unit volume is given by
$$U_{in}=\sum_iU_{in}^{(i)}\eqno(6.22)$$
where
$$U_{in}^{(i)} =\eta_iB_m \left (n_i^{(\uparrow)}(B_m)
-n_i^{(\downarrow)}(B_m) \right )\eqno(6.23)$$
Substituting the values of $n_i^{(\uparrow)}$ and $n_i^{(\downarrow)}$
and after some simple algebraic manipulation, we have
$$U_{in}^{(i)}= n_i \eta_i  B_m  \left  [1-
{{x_i^{3/2}}\over{ R_{3/2}(x_i)}} \right ]\eqno(6.24)$$
which becomes zero for $B_m=0$ and at the saturation limit,
$$U_{in}^{(i)}(x_i=1)=n_i \eta_i B_m\eqno(6.25)$$

Finally,  the  chemical potential for the species $i$ in presence
of a magnetic field $B_m$ is given by
$$\mu_i=\epsilon_i -\eta_i B_m\eqno(6.26)$$
Substituting the values for $\epsilon_i$ and $\eta_iB_m$, we have
after some simple algebra
$$\mu_i=\left \{ m_i^2  +x_i[R_{3/2}]^{-2/3}  (\epsilon_{(i,0)}^2
-m_i^2)   \right   \}^{1/2}   -{{1}\over{2}}  \left  \{  \left  [
{{\epsilon_i^2-m_i^2(1- x_i^2)}\over{x_i}}  \right  ]^{1/2}  -m_i
\right \} \eqno(6.27)$$
which  becomes  $\epsilon_{i,0}$  for  $y_i=0$,  or  equivalently
for $B_m=0$ or $x_i  \rightarrow  \infty$.  On  the  other  hand  in the
saturation limit, we have
$$\mu_i(x_i=1)={{1}\over{2}}  \left [ \epsilon_i(x_i=1) +m_i \right
]\eqno(6.28)$$

\medskip
\noindent {\bf{7. ~Conclusions and Discussions}}
\medskip

From  the theoretical investigations of SQM in presence of strong
magnetic field we can make the following conclusions:

(i)  In presence of a strong magnetic field  of  strength  of the
order of greater than some critical value  as  discussed  in  the
main  text,  the SQM becomes energetically more stable as long as
$B_m\leq 10^{20}$G. Above this value since  the  magnetic  energy
density  and  also  the magnetic pressure play the main roles the
system becomes energetically as  well  as  mechanically  unstable
(pressure of the system becomes negative).

(ii)  Equation  of state of magnetized SQM changes significantly.
The $\beta$ equilibrium point has also been shifted.

(iii) There can not be any thermal nucleation of  quark  droplets
in a metastable state of compact neutron matter.

(iv)  If bulk QGP and hadronic matter can co-exist in presence of
strong magnetic field, then a significant change in phase diagram
has also been  observed.

(v)  If  the  quark-hadron phase transition at low temperature
and high density is
like a metal insulator type of second order transition.  Then  in
presence  of extremely  strong  magnetic  field  as  mentioned  above,  the
critical density at which such transition may occur is  found  to
be  too large, which is almost impossible to achieve at the
core of a neutron star.
\vfil\eject
\noindent {\bf {References}}
\item{1.} E. Witten, Phys. Rev. D30, 272 (1984).
\item{2.}  S. Chakrabarty, S. Raha and B. Sinha, Phys. Lett. 229B, 112
(1989).
\item{3.} E. Farhi and R. L. Jaffe, Phys. Rev. D30, 2379 (1984).
\item{4.} P. Haensel, J. D.  Zdunik  and  R.  Schaeffer,  Astron.
Astrophys. 160, 12 (1986).
\item{5.} C. Alcock, E. Farhi and A. V. Olinto, Ap. J. 310, 261 (1986).
\item{6.} H. Reinhardt and B.V. Dang,  Phys.  Lett.  202B, 133  (1988);
 T. Chmaj and W. Slominski, Phys. Rev. D40, 165 (1989).
\item{7.} S.  Chakrabarty, Phys. Scr. 43, 11 (1991).
\item{8.}  S. Chakrabarty, Phys. Rev. D43, 627 (1991).
\item{9.} S. Chakrabarty, Mod. Phys. Lett. A9, 187 (1994);
S. Chakrabarty, Mod. Phys. Lett. A9, 2691 (1994).
\item{10.}  C.  Alcock,  Numerical  Experiment  and  Neutron Star
Formation, Proceedings of IAU Symposium,  Nanjing,  China,  1986,
ed. D. J. Helfand  and  J.  -H.  Huang,  IAU  Symposium  No.  125
(Reidel, Dordrecht, 1987), P. 413.
\item{11.}  F.  Grasi, Z. Phys. C44, 129 (1989); A. Rosenhauer and
E.F. Staubo, Nucl. Phys. (Proc. Suppl.) 24B, 156 (1991).
\item{12.} N. K. Glendenning, Phys. Rev. Lett. 63, 2629 (1989);  J.
Madsen, Phys. Rev. Lett. 61, 2902 (1988).
\item{13.}  J.  E. Horvath, O.G. Benvenuto and H. Vucetich, Phys.
Rev. D45, 3865 (1992); Univ.  of  Sau  Paulo  Preprint,  1994  (to
appear in Phys. Rev. D).
\item{14.}  See e.g. Black Holes, White Dwarfs and Neutron Stars,
The Physics of Compact Objects, S.L. Shapiro and  S.A.  Teukolsky
(Wiley, New York, 1983).
\item{15.}  D.  Persson  and  V. Zeitlin, Preprint FIAN/TD/94-01,
hep-ph/9404216; V. Zeitlin, FIAN/TD/94-10.
\item{16.} A.S. Vshivtsev and D.V. Serebryakova  JETP  79, 17  (1994).
\item{17.}   Ulf   H.   Danielsson  and  Dario  Grasso,  Phys. Rev
D52, 2533 (1995).
\item{18.} B. Cheng, D. N. Schramm and J. W. Truran, Phys.  Lett.
B316, 521 (1993); B. Cheng, D. N. Schramm and J. W. Truran, Phys.
Rev. D49, 5006 (1994).
\item{19.}  D. Graso and H. R. Rubinstein, Astropart. Phys. 3, 95 (1995);
Preprint UUITP-3/96.
\item{20.} S. Chakrabarty, Astron. Space Sci. 213, 121 (1994).
\item{21.} S. Chakrabarty and A.K. Goyal,  Mod.  Phys.  Lett.  A9, 3611
(1994).
\item{22.} S. Chakrabarty, Phys. Rev. D51, 4591 (1995).
\item{23.}   S.   Chakrabarty   Jour.  Astronomy  \&  Astrophys.
(in press);  S.  Chakrabarty    Phys.  Rev.  D
(submitted); Astrophys. and Space Sci. (Submitted).
\item{24.} S. Chakrabarty, Can J. Phys. 71, 488 (1993); S. Chakrabarty and
P.K. Sahu, Phys. Rev. D (in press).
\item{25.}  L.  D.  Landau  and E. M. Lifshitz, Quantum Mechanics
(pergamon Press, 1965).
\item{26.}  L.  D.  Landau  and  E.  M.   Lifshitz,   Statistical
Mechanics, Part I, Vol. 5 (Pergamon Press, 1980).
\item{27.}  M.S.  Berger,  Phys.  Rev. D40, 2128 (1989); M.S.  Berger,
Phys.  Rev. 43, 4150E (1991).
\item{28.} J. Madsen, Phys. Rev. D50, 3328 (1994);  D.  M.  Jensen
and J. Madsen, Preprint, Aarhus-Astro-1994-18.
\item{29.} P. K. Sahu and S. Chakrabarty (in preparation).
\item{30.}D. Turnball and J.C. Fisher, Jour. Chem. Phys. 17, 71 (1949);
W. Kurz and D. J. Fisher, Fundamentals of Solidification, (Trans Tech Pub.,
USA, 1984); see also, S. Chakrabarty, J. Phys. G20, 469 (1994).
\item{31.} M.L. Olesen and J. Madsen Phys. Rev. D49, 2698 (1994).
\item{32.}  L.P.  Csernai  and J. I. Kapusta, Phys. Rev. Lett. 69, 737
(1992); L.P.  Csernai  and J. I. Kapusta, Phys. Rev. D46, 1379 (1992).
\item{33.} C. Kittel, Introduction to Solid State  Physics  (John
Wiley and Sons, 1986).
\vfil\eject
\noindent {\bf{Figure Captions}}
\item{1.}  The  variation  of  $u$,  $d$, $s$-quarks and electron
number   density   with   the  magnetic  field  intensity  $B_m$,
$n_B=2.5n_0$.
\item{2.} Plot of energy per baryon  against  the  baryon  number
density   for   $B_m=0$   (upper   curve)   and   $B_m=10^3\times
B_m^{(c)(e)}$ (lower curve).
\item{3.} Pressure vs  energy  per  baryon  curves  (equation  of
states)  for  $T=0$. The dotted curve is for $B_m=0$ and solid one
is for $B_m=10^3\times  B_m^{(c)(e)}$.  The  curve  indicated  by
$bj$ is for B-J equation of state.
\item{4.}  Pressure  vs  energy  per  baryon  curves (equation of
states) for $T=40$MeV. The dotted curve is for $B_m=0$ and  solid
one is for $B_m=10^3\times B_m^{(c)(e)}$.
\item{5.}  variation of energy per baryon for stable SQM with the
magnetic field strength for $n_B=2.5n_0$ and $T=0$.
\item{6.}  variation of pressure with the
magnetic field strength for $n_B=2.5n_0$, $T=0$ and $B^{1/4}=0$.
\item{7.} Phase diagram for a bulk co-existing QGP  and  hadronic
phases   for   $B_m=0$   (dotted   curve)   and   $B_m=10^3\times
B_m^{(c)(e)}$ (solid curve).
\item{8.} variation of $n_c$ with the  magnetic  field  intensity
$B_m$.
\item{9.} variation of $B^{1/4}$ with the  magnetic  field  intensity
$B_m$.
\vfil\eject\end